# Single photon sources: ubiquitous tools in quantum information processing

Urbasi Sinha[*], Surya Narayan Sahoo, Ashutosh Singh, Kaushik Joarder, Rishab Chatterjee, Sanchari Chakraborti

Light and Matter Physics Group, Raman Research Institute, Sadashivanagar, Bangalore-560080, India.

[*]Corresponding author: usinha@rri.res.in

**Abstract**: Quantum technologies are the next big revolution in information technologies, computing, communication security, sensing as well as metrology. What do you use to explore all these fascinating applications when you work in Optics? Photons of course. In this review, we discuss the different available single photon source technologies, compare and contrast them in terms of applicability and properties, discuss state of the art and conclude that the future is indeed bright!

Quantum Technologies can revolutionise security in communications, computational speed as well as high precision sensing. The saying goes "Just like the 19[th] century was called the machine age, the 20[th] century the information age, the 21[st] century is the quantum age". Indeed, the last few decades have seen remarkable developments in applications of quantum physics to problems of fundamental as well as technological relevance.

While there are many candidate systems for developing and testing fundamental as well as technically innovative ideas in the quantum domain, optical test beds have always remained a popular choice for first tests of many such schemes. This is primarily due to the inherent simplicity and beauty of optical demonstrations as well as in general lower cost requirements than for instance many solid state systems which are prohibitively expensive due to manufacturing costs as well as measurement conditions often requiring cryogenics.

When we talk of optical implementations of quantum technologies, the "quantum-ness" can be brought in to begin with, at the detection level. Herein, by using photon counting techniques, we can conceive of a particle like behaviour for the incident radiation. However, true quantum-ness can usually be manifested when the source itself is also quantum. In optics, that would mean the single particle of light which is called a photon.

How do you generate a single photon?

One might say, just take some strong light source, then keep on attenuating it such that you reach a level where the mean number of photons in a given time frame reaches a number ~ 1 or less than 1. This is what is popularly called an attenuated laser source or weak coherent pulse (WCP). While this is indeed a popular choice, it suffers from not being truly quantum in the sense that it exhibits different statistics from what we would expect a single photon source to follow.

In this article, we discuss different types of currently available photon sources [1-10], compare and contrast them in terms of their performance matrix vis-à-vis various source properties and also in terms of their suitability towards applications in various quantum technologies (see Table [1]). As we go along we also introduce a few fundamental quantum optics concepts which are required for such an overview.

An important thing to consider when we talk about single photon sources is "How readily available are the photons?" Now this may seem like a strange question but in photon source technology, it is actually a very important consideration. Can they be available on demand or is the availability probabilistic? While many things can be done using probabilistic sources, some applications would prefer the on-demand version.

Are the photons "entangled" i.e. do they share the special quantum correlation which makes them useful in so many information processing and communication protocols? While some photon generation schemes lend themselves towards generating Entanglement more easily, some others are more tedious to entangle [2].

How bright is the source? For applications like generating secure keys in Quantum Communication over long distances, brightness plays an important role. As the photons travel over long distances, they suffer losses due to various factors and the resultant key rate is thus high only when the original source had a relatively large number of photons being emitted per unit time.

While we talk of photons, we need to care about their "colour". In other words, the wavelength of emission as well as its attendant linewidth matter. For instance, for communication purposes, telecommunication wavelengths have an advantage when the medium is optical fibre based. In free space on the other hand, we look for windows where scattering is less and transmission is higher which cover near infra-red, telecom as well as other bands.

What are photons most useful for? Well, photons are chargeless, massless particles, interact very little with neighbouring photons and are easily manipulated and detected. Being thus isolated beings, they can literally go very far i.e. they are perfect candidates for being the carrier in communication technologies. On the other hand, when we talk of quantum computing, we need to build quantum gates. While for instance the polarization degree of freedom of a single photon leads to a good single qubit and suitable single qubit gates, quantum gates are also needed to couple multiple qubits. This does not naturally happen with photons as they don't naturally interact. However, there are ways to make this happen, popularly by mediating this interaction through the presence of matter. Such light-matter interaction leads to realizations of multiple qubit gates. This can also happen through other indirect processes like heralded gates etc. A single photon is a popular tool for applications in quantum metrology including quantum sensing among others. That being said, no other technology can perhaps compete with photons for their usefulness in being perfect test beds for testing various foundational ideas and principles in quantum mechanics. Most of the seminal experiments in quantum foundations (many of which have intricate implications on quantum technologies) have been performed first using photons, paving the way for other implementations using atoms, ions and molecules. These include but are not limited to the violation of Bell inequality and various loophole free versions of the same, other tests of non-locality, superdense coding, teleportation, entanglement swapping, steering, tests of weak interactions and measurements, quantum cryptography among others.

How do we know that we indeed have produced a single photon? In other words, what are the popular photon source characterization techniques? One of them is called the $g^{(2)}$ measurement. $g^{(2)}$ is a second order correlation function, which in this context involves a single photon incident on one of the input ports of a 50/50 beam splitter and the cross correlation function being measured between the two detector outputs (see Fig. [1]). As the input is a single photon, it can exit from either of the output ports but of course not through both as a single photon does not split into two. Thus, there is no cross correlation between the two detector outputs. This is what leads to a dip at zero time delay and gradually non-zero values for the cross-correlation as a function of time delay between the two detections. The value of $g^{(2)}$ for zero time delay or $g^{(2)}(0)$ being as close to zero as possible is a signature of a true single photon source. For a comparison, $g^{(2)}(0)$ for a laser source is one.

Another common characterization tool for photon sources also involves the beam splitter. But, in this case, we have two completely indistinguishable single photons incident on the two input ports (one in each input port) of a 50/50 beam splitter (see Fig. [1]). Due to interference effects, only the possibility of both exiting photons reaching the same detector survives and one photon each going to individual detectors cancels out. This leads to the formation of what is called a superposition state between the two 2-photon possibilities i.e. $\frac{1}{\sqrt{2}}(|2,0\rangle - |0,2\rangle)$. Thus, when a cross correlation measurement is performed between the two output ports of the beam splitter, there is no correlation captured at zero time delay between the two input photons with the curve gradually allowing non-zero correlation values with increasing time delay between the inputs. This is called the Hong-Ou-Mandel (HOM) effect. While both the $g^{(2)}$ and the HOM involve cross correlation measurements between beam splitter outputs, the $g^{(2)}$ only captures the single photon character of the source while the HOM also characterizes the distinguishability between the input photons.

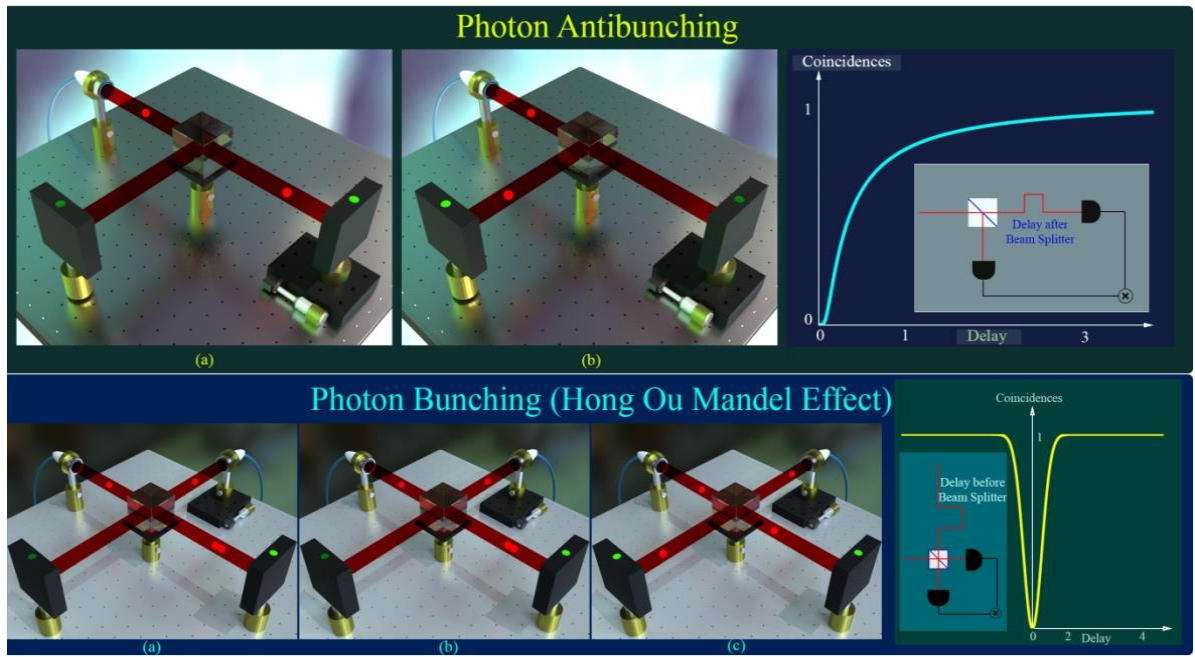

**Fig. (1): Anti bunching($g^{(2)}$):** A single photon is an indivisible entity. Thus, when it is incident on a beam splitter, it either gets transmitted (a) or gets reflected (b). Thus both the detectors do not click simultaneously and hence at zero delay of the output signals there is no correlation. However, the correlation increases as we increase the delay. **Photon Bunching:** When two indistinguishable photons are incident on two ports of a beam splitter they always exit together at the same output port (a) and (b). However, if there is a distinguishability factor, say a slight delay in the arrival time, there is chance that each of them gets reflected or transmitted thus exiting at different ports(c). The correlation of the detectors is thus zero for zero delay in arrival time of photon but increases if delay is increased.

Now we go on to discussing different types of currently available schemes of generating single photons. We begin this discussion with the most widely used and popular means of generating single photons which is called Spontaneous Parametric DownConversion (SPDC) [8,10]. This is a second order non-linear optical process in which a high energy pump photon, in the presence of a non-linear medium, spontaneously generates two lower energy photons; historically known as "signal" and "idler" photons (see Fig.[2]). The process is spontaneous (generated by quantum vacuum fields), parametric (initial and final quantum mechanical states of the medium are identical and photon energy is always conserved) and an example of down-conversion in the sense that the signal and idler frequencies are always lower than that of the pump. The SPDC process follows energy and momentum conservation which are together known as the "phase matching conditions". The SPDC process can be used in the heralded scheme to generate single photons. As both photons are generated at the same time, detecting one heralds the presence of the other and this stream of heralded photons can then be used for single photon applications. So, what do we gain and what do we lose when we use SPDC?

There is of course a lot to gain. In spite of the efficiency of the process being low (1 in $10^9$ or 1 in $10^{10}$ pump photons undergo the SPDC process) but still SPDC process generates some of the brightest photon sources (one of the brightest being ~2 MHz). One of downsides of the SPDC process is the fact that it is probabilistic. Thus, if an application requires a photon on-demand, SPDC is not our solution. Moreover, there is a finite probability of multi-photon events which can cause security loopholes in quantum key distribution protocols; although the situation is much better than weak coherent pulses where such probabilities are much higher.

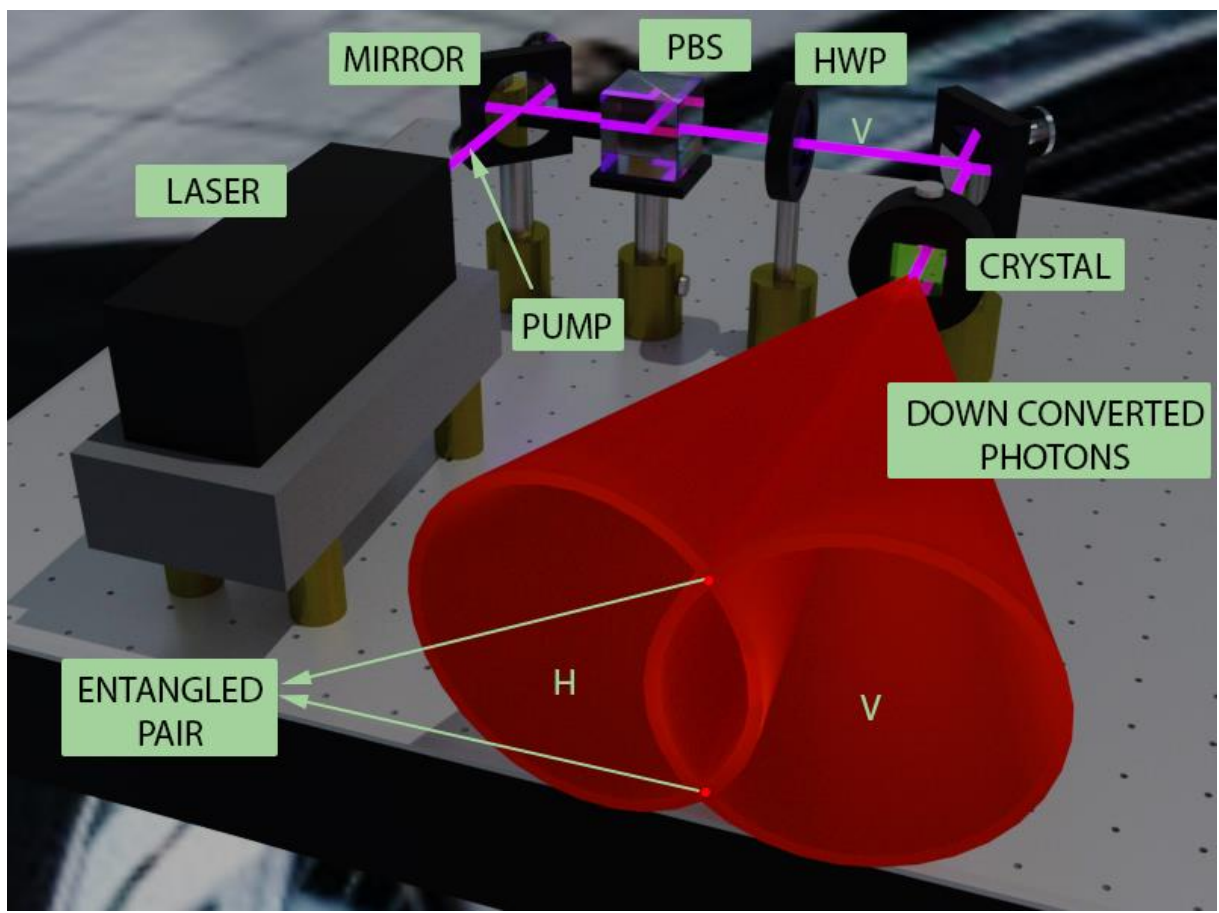

**Fig. (2):** A pump laser (UV/IR) in continuous wave or pulsed mode is incident on a non-linear medium which is a birefringent crystal. The crystal facilitates the process of spontaneous parametric down conversion in which one pump photon is absorbed to generate two daughter photons. The distribution of the generated photons satisfies the phase matching conditions, which in a Type-II crystal gives the shape of two displaced cones. The polarization of the pump can be adjusted such that one of the emergent cones has vertically polarized photons while the other has horizontally polarized photons. In the region, where the two cones intersect, each point can have either a Horizontal or a Vertically polarised photon a priori with equal probability. But if one is detected to be Horizontal, the type-II process ensures that the other is Vertical, thus making the photons collected from the intersection regions entangled.

What about generating entangled photons? SPDC clearly leads the way. Greater than 99.5% fidelity of entanglement has been measured in entangled photons produced by SPDC. Thus this process finds numerous applications in foundational tests of quantum mechanics, quantum information processing, quantum metrology as well as quantum communication.

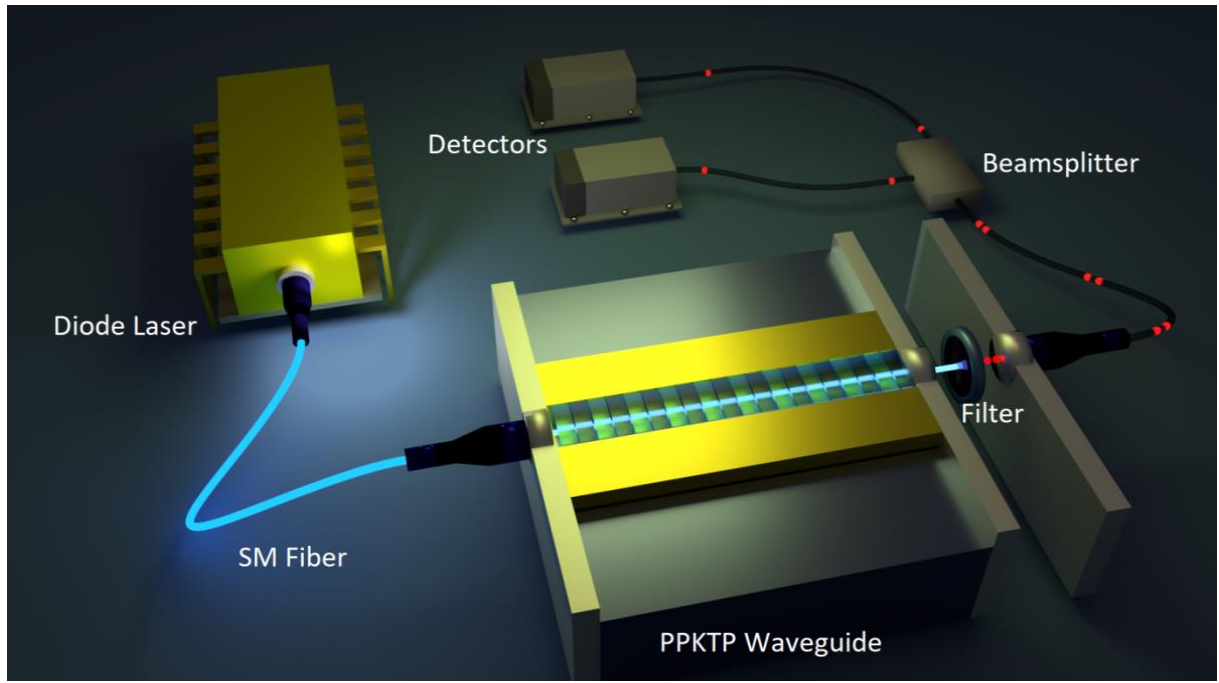

**Fig. (3)**: A continuous wave (CW) pump beam is coupled to a single mode fibre. This fibre is properly aligned to a PPKTP waveguide, which also maintains the Gaussian spatial mode of the pump beam. Due to SPDC process photon pairs are generated continuously (red bullets in the figure) and passed through a filter which blocks any pump light. Two photons whose polarizations are orthogonal to each other, can be separated by a polarizing beam-splitter and detected in two single photon avalanche detectors (SPAD).

Another scheme which uses a non-linear medium to generate photons is called four-wave mixing [10]. Here, two pump photons convert to a signal and an idler photon in the presence of a non-linear medium. This also requires satisfaction of phase matching conditions. As opposed to SPDC which can occur both in bulk as well as in confined geometries like cavities as well as waveguides (which in general narrows down the emission direction), four-wave mixing has been generally demonstrated only in integrated optics structures involving waveguides. Waveguide based sources in general can have higher pair generation rate compared to bulk sources due to less number of interacting modes (see Fig.[3]). Four wave mixing based photon sources have reported brightness of ~0.855 MHz. The current record for entanglement fidelity is ~99.7%.

Till now, we have been dealing with probabilistic sources. Now, we go on to discussing some on-demand options for photon sources. One of the simplest and most elegant architectures for a single photon source involves emission of photons due to transitions in atoms or ions [10]. Historically, the first so called single photon source was an atomic cascade based source. Seminal experiments in quantum foundations like demonstration of Bell inequality violation as well as double slit interference using photons were performed using atomic cascade sources. A single excited atom spontaneously emits a single photon but isolating the single atom was a challenge. The first anti-bunching experiment was successfully observed in resonant fluorescence of sodium atoms. Then atomic trapping was invented which led to the investigation of trapped ion as well as trapped atom based photon sources. Emission from such sources happens in all directions which leads to poor detection efficiency. A way to increase this would be to place atoms/ions in a cavity. Then the emission modes couple to the cavity mode giving it directionality, thus increasing the efficiency. The important advantage of trapped atom/ion sources is the narrow emission line width they offer. Thus, they are very good in interferometric applications. They are pure single photons as the probability of multiple photon emission is quite low in general. While not the optimal choice for technological applications, they remain an excellent choice for experiments in quantum foundations as well as metrology.

Now that we talked about atoms and ions, what about "artificial atoms" a.k.a quantum dots? Indeed, one of the most promising upcoming photon sources with a lot of potential for future developments and applications [9,10]. Quantum dots are tiny particles or nanocrystals of a semiconductor material with diameter in the range of 2 – 10 nm. They are referred to as artificial atoms owing to their atom-like discrete energy spectra. When illuminated by an optical pulse, electrons in the quantum dot and in the vicinity jump from valence to conduction band, thus leading to the formation of electron-hole pairs which then rapidly relaxes to the lowest lying energy states (see Fig. [4]). Depending on the population, one can observe recombination from the exciton, the bi-exciton and multi-exciton state which refers to the number of electron-hole pairs left to recombine. Although all such transitions possess distinct energies and can in principle be used as an emitter for single photons by spectral filtering, the bi-excitonic and the excitonic transitions are the most commonly used [9].

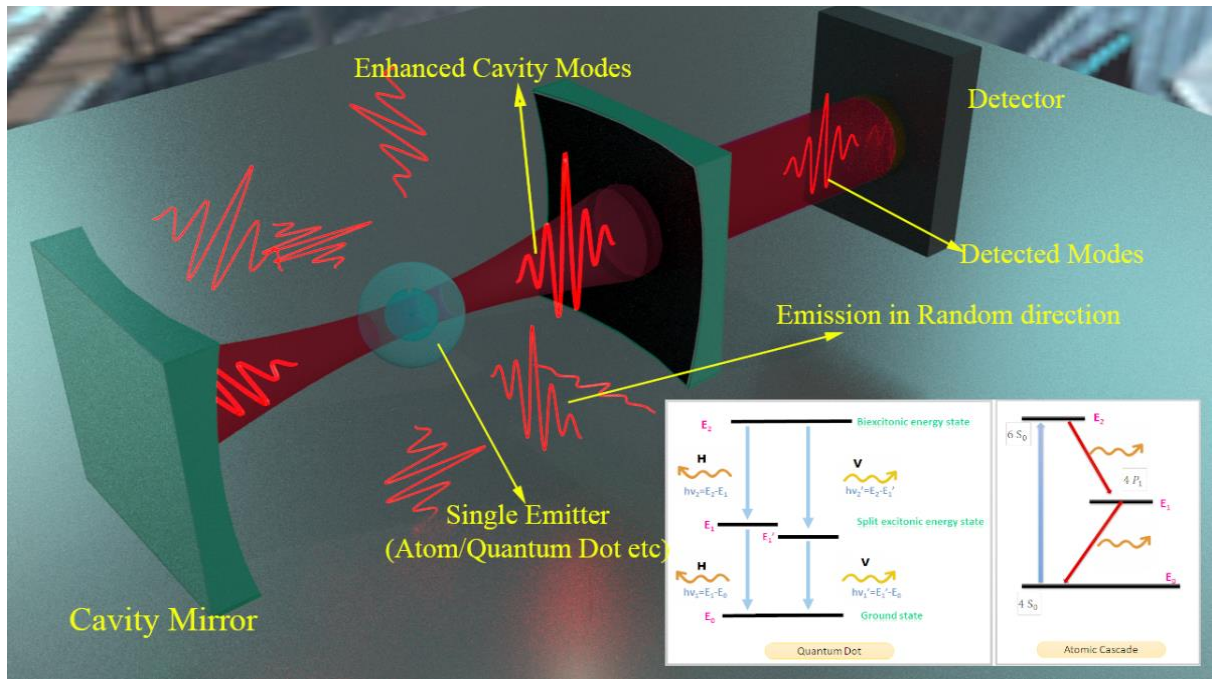

**Fig. (4)**: Atoms or artificial atoms like Quantum dots when excited undergo spontaneous emission of photons in all directions. Atomic cascades and Bi-exciton modes in Quantum Dots have small lifetime of metastable states and thus emit photons in a pair. When we place atoms/quantum dots in a cavity, the emission modes resonant to the cavity get enhanced and amplify the emission in a given spatial direction.

Quantum dot sources have also been used to produce highly entangled photon pairs. While the current highest single photon emission rate stands at 28.3 MHz, entanglement fidelity of 0.978 has been demonstrated. A quantum dot source can be either optically pumped or electrically pumped and while the electrical pumping technique holds the higher brightness record, the $g^{(2)}(0)$ is quite high for the same compared to the optically pumped ones. With electrical pumping, sources have been realized even at room temperature. Self-assembled quantum dots have been integrated with cavities to increase directionality of emission as well as ease of use and the structural integrity of these cavities continues to improve thereby enhancing the quality factor. For epitaxial quantum dots, the dots sometimes grow at random positions on a surface which leads to losses in collection efficiency. Continuous improvements are happening in this domain with brightness and entanglement records being broken frequently. These deterministic sources have been successfully used in various quantum key distribution implementations as well as experimental demonstration of multi-photon Boson sampling among other applications.

| Source | P/D | Emission Range | Band-width | Operating Temperature | Emission Direction | Efficiency | Max. Brightness | Best $g^{(2)}$ | Entanglement Fidelity | HOM Visibility | Application[†] |
|---|---|---|---|---|---|---|---|---|---|---|---|
| SPDC | P | 600-1700 nm | nm | 0-200 °C | Narrow | 0.84 | 2.01 MHz[a] | 0.004 | 0.9959 | 0.99 | QM, QI QF, QC |
| Atoms & Ions | P,D[b] | Transition Lines | 10 MHz | Room Temp, mK (in cavity) | Random, Narrow[c] | 0.88 | 55 kHz | 0.0003 | 0.93 | 0.93[d] | QF |
| Quantum Dots | D | IR, Telecom | nm | Room Temp., Cryogenic[e] | Random, Narrow | 0.97 | 28.3 MHz | 0.000075 | 0.978 | 0.9956 | QC, QF |
| NV Centre | D | 600-800 nm | 1-100 nm[f] | 300-500 K | Random | 0.35 | 850 kHz[g] | 0.07 | ---[h] | 0.66 | QC, QN |
| 4 Wave Mixing | P | 600-1550 nm | 10 nm | Room Temp. | Narrow[b] | 0.26 | 855 kHz | 0.007 | 0.997 | 0.97 | IP |

**Table (1):Comparison of properties and yield of various types of single photon sources**
All the parameters mentioned above for a given source type do not necessarily belong to the same study. Some properties like high brightness and maximum anti-bunching may not be achievable simultaneously.

a) The highest observed spectral efficiency is 0.41 MHz/mW/nm but here we have reported the highest raw counts or coincidences.
b) Deterministic in a cavity, but probabilistic in general.
c) Cavity makes the emission direction narrow.
d) The visibility is 0.93 after background correction. The raw value is 62%.
e) Cryogenic for Colloidal Quantum Dots and Room Temperature for Epitaxial Quantum Dots.
f) It's about 1 nm for the ZPL and 100 nm for the broadband emission.
g) Count rates of 35 MHz have been achieved with NV centre but they have $g^{(2)} \gg 0.2$ (850 KHz brightness gives 0.08 $g^{(2)}$ )
h) While entanglement between NV centre with another NV centre or photon has been demonstrated, generation of entangled pair from NV centre is currently at the stage of theoretical proposal.

[†]QM=Quantum Metrology, QI= Quantum Information, QF = Quantum Foundations, QC = Quantum Communication, QN = Quantum Network, Integrated Photonics

P/D: Probabilistic source/Deterministic source

Efficiency: For SPDC and FWM, heralding efficiency is considered whereas for atoms and ion sources the efficiency with which photons are generated in cavity is mentioned. Usually the collection efficiency for non-cavity based sources would make the overall efficiency as low as 0.0001. For Quantum Dots and NV Centre we report the Quantum Efficiency.

Entanglement Fidelity: It is a measure of overlap of the obtained state with the expected entangled state.

Maximum Brightness: For SPDC, we report the raw heralded coincidence counts that are detected. For atom/ion source we report the fluorescence observed. Further, chip based waveguide sources have been demonstrated to have high count rates of 2.1 MHz with $g^{(2)}$ of 0.023.Furthermore, with single molecules, detection rates of 310 KHz have been achieved.

Another upcoming, promising technology is based on colour centres in carbon. In a diamond structure, missing carbon atoms lead to defects. Instead of replacing these defects with carbon, one may be replaced with a nitrogen atom while the other defect could be left vacant. This is known as a Nitrogen-Vacancy (NV) centre in diamond. Similarly, instead of nitrogen, other atoms can also be used to replace the carbon which leads to a family of diamond colour centres.

NV centres are promising candidates for single photon generation and since the emission rate has been seen to increase with temperature, they are potential candidates for room temperature on demand photon sources [10].

With all the discussion above, it is quite clear that a lot of work has gone into realizing and trying to perfect single photon sources. What then makes photon source technology and its associated research still so interesting? We have established by now that photons by themselves are ubiquitous in the various technological leaps that we see around us in terms of communication, metrology as well as in computing. We also know that there are different ways of creating these photons. What next? Well, we have found that each of the techniques that has been discussed so far has its attendant scope of improvement. While some are more established in their usage and are at near optimal performance, some others hold more scope for further research and new records to be broken. And while we go on working towards improving those numbers, making sources with higher brightness, higher entanglement and greater stability, efficiency and indistinguishability; we look forward to a future world of the quantum internet where the entire communication network is replaced by a quantum version with satellites transmitting photons to the earth and breaking distance barriers; optical fibres connecting far away nodes to form backbone networks and quantum memory research becoming more advanced in storing the much valued information. We also cannot forget the role of the photon in quantum computing both as a carrier in hybrid schemes as well as the basis for linear optics based quantum computing. Last but not the least, while we go on making advances with far reaching technologies, quantum mechanics and quantum optics continue to fascinate us with new intrigues thrown in every now and then which need to be tested and verified through precision experiments and what can be a better candidate for such tests than a single photon?

## **Further Reading:**

1) *"Experimental Methods in the Physical Sciences", Chapter 10-14, Editors : Alan Migdall et.al , Academic Press, Volume 45, 2013, Pages 315-562, ISBN 9780123876959.*
2) *S Takeuchi, "Recent progress in single-photon and entangled-photon generation and applications",Jpn. J. Appl. Phys.* **53***, 030101 (2014).*
3) *M. D. Eisaman, J. Fan, A. Migdall and S. V. Polyakov, "Invited Review Article: Single-photon sources and detectors", Rev. Sci. Instrum.* **82***, 071101 (2011).*
4) *B. Lounis and M. Orrit, "Single-photon sources", Rep. Prog. Phys.* **65,***1129 (2005).*
5) *M. Oxborrow and A Sinclair, "Single-photon sources", Contemp. Phys.* **46***, 173 (2007).*
6) *S. Scheel, "Single-photon sources–an introduction", J. Mod. Opt.* **56***, 141 (2010).*